\documentstyle[times,balanced,epsf,prl,aps]{revtex}

\begin{document}

\draft

\title{Burridge--Knopoff Models as Elastic Excitable Media}

\author{
Julyan H. E. Cartwright$^{1,2,}$\cite{jemail}, 
Emilio Hern\'andez-Garc\'{\i}a$^{1,3,}$\cite{hemail}, and 
Oreste Piro$^{1,3,}$\cite{oemail}
} 

\address{
$^1$Departament de F\'\i sica, Universitat de les Illes Balears,
07071 Palma de Mallorca, Spain \\ 
$^2$Centre de C\`alcul i Informatitzaci\'o, Universitat de les Illes Balears,
07071 Palma de Mallorca, Spain \\ 
$^3$Institut Mediterrani d'Estudis Avan\c{c}ats, IMEDEA (CSIC--UIB),
07071 Palma de Mallorca, Spain \\ 
}

\date{Physical Review Letters, {\bf 79}, 527--30, 1997}

\maketitle

\begin{abstract}
We construct a model of an excitable medium with elastic rather than the usual
diffusive coupling. We explore the dynamics of elastic excitable media,
which we find to be dominated by low dimensional structures, including global 
oscillations, period-doubled pacemakers, and propagating fronts. We suggest 
that examples of elastic excitable media are to be found in such diverse 
physical systems as Burridge--Knopoff models of frictional sliding, 
electronic transmission lines, and active optical waveguides.
\end{abstract}


\begin{twocolumns}

The response to a perturbation is the defining characteristic of an excitable 
element. Above a certain threshold in amplitude, a perturbation will excite a 
quiescent element that then decays back to quiescence after a characteristic 
time insensitive to the magnitude of the perturbation, during which it is
unresponsive to further perturbations. Excitable elements, when coupled to 
their neighbors into an assembly, become an excitable medium. Models of 
excitable media with diffusive coupling, such as those of van der Pol, 
FitzHugh, and Nagumo \cite{vdpfn} have successfully described pattern 
formation phenomena in biology and chemistry, and have also captured the 
attention of many working in the area of nonlinear science because of their 
complex spatiotemporal dynamics \cite{meron}. On the other hand, the 
properties of excitable media with elastic rather than diffusive coupling 
have not to date been investigated. We present here some physical systems --- 
frictional sliding, electronic transmission lines, and active optical 
waveguides --- as diverse examples of such elastic excitable media, and analyze 
their dynamics, which we find to be dominated by low dimensional structures, 
including global oscillations, period-doubled pacemakers, and propagating 
fronts. We investigate the stability of the oscillations, and estimate the 
velocity of the fronts.

Whilst touching on electronic and optical applications of 
elastic excitable media, we shall concentrate on frictional sliding. 
The Burridge--Knopoff model \cite{burridge} was originally introduced as a 
representation of earthquake fault dynamics. It describes the interaction of 
two tectonic plates in a geological fault as a chain of blocks elastically 
coupled together and to one of the plates, and subject to a friction force by 
the surface of the other plate, such that they perform stick--slip motion. 
In this Letter, we show that a Burridge--Knopoff model in which the usual 
Carlson--Langer velocity weakening dry-friction law \cite{carlson} is replaced 
by the original Burridge--Knopoff lubricated creep--slip version showing 
viscous properties at both the low and high velocity limits, is an elastic 
excitable medium.

The reintroduction of this friction law in the Burridge--Knopoff model is 
motivated by experiments showing that it represents friction in a range of 
materials, including paper on paper \cite{heslot}, metal on metal 
\cite{brechet}, and rock on rock \cite{kilgore}, and by theoretical arguments 
\cite{persson}. The model well represents the qualitative characteristics of 
laboratory stick--slip experiments with elastic gels \cite{rubio}. Further 
interest derives from studies of peeling adhesive tape \cite{hong}, of 
Saffman--Taylor fracture in viscous fingering \cite{kurtze}, and of the 
Portevin--Le Ch\^atelier effect of discontinuous yielding \cite{kubleb}, 
showing that stick--slip phenomena in those systems are induced by this same 
form of friction law. 

Burridge \& Knopoff introduced a class of 
simple models that describe the contact region between two tectonic plates 
as a chain of $N$ blocks of equal mass $m$, mutually coupled by springs of 
Hooke constant $k_c$ and equilibrium length $a$. The blocks are pulled by 
the bulk of one plate moving at velocity $V$ through constant elastic shear 
$k_p$ against the friction $F_f$ between the two plates, as 
shown in Fig.\ \ref{friction}(a). In the stationary frame, the equation of 
motion for the $i$th block is
\begin{eqnarray}
m \ddot X_i=k_c(X_{i+1}-2 X_i+X_{i-1}) \nonumber \\
\mbox{}-k_{p}(X_i-Vt)-F_f(\dot X_i),
\label{eom}\end{eqnarray}
where $X_i$ is the departure of block $i$ from its equilibrium position. 
Usually, following Carlson \& Langer \cite{carlson}, the friction is taken 
to be asymptotically velocity weakening, so that the blocks perform 
stick--slip motion: any individual block sticks to the surface until the 
pulling force exceeds the static friction threshold. Once the block starts 
slipping, the dynamic friction diminishes monotonically with its velocity, as
Fig.\ \ref{friction}(b) illustrates. While this multivalued form of the 
friction law is perfectly admissible in the discrete model of 
Eq.\ (\ref{eom}), it clearly poses a problem if we attempt to take the 
continuum limit \cite{carlan}. 

\begin{figure}
\begin{center}
\def\epsfsize#1#2{0.46\textwidth}
\leavevmode
\epsffile{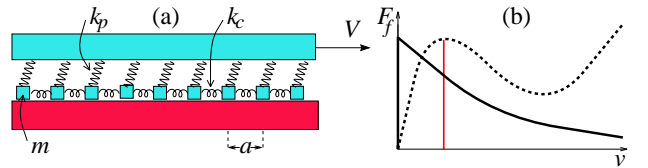}
\end{center}
\caption{\label{friction} (a) The Burridge--Knopoff model. 
(b) The velocity weakening stick--slip friction law of Carlson \& Langer
[$F_f(v)=F_0\,{\mathrm sgn}(v)/(1+|v|)$, where $v$ is the velocity of 
the block] (solid line) and the Burridge--Knopoff type creep--slip friction 
law we use (dashed line), showing the threshold (vertical line) where the block 
starts to slip.
}
\end{figure}

To overcome this difficulty, one may replace the static friction discontinuity
by a small region of high viscosity.  Physically, one can see this cut-off as 
representing lubrication effects that produce stable creep at low velocities,
making our model creep--slip rather than stick--slip. Since we wish to model 
the frictional characteristics displayed by a range of materials 
\cite{heslot,brechet,kilgore}, we also have the friction become viscous 
(velocity strengthening) at high velocities, rather than decay to zero as in 
the Carlson--Langer version. Our friction model (Fig.\ \ref{friction}(b)), 
which is similar to that originally introduced by Burridge \& Knopoff,
thus comprises three regions: first velocity strengthening, then 
velocity weakening, and finally velocity strengthening again. We shall 
refer to this model as {\it asymptotically velocity 
strengthening\/}, to be contrasted with the {\it asymptotically velocity 
weakening\/} friction usually considered. With this in mind, the 
continuum limit of Eq.\ (\ref{eom}) in dimensionless variables 
(see, e.g., \cite{carlan}) is 
\begin{equation}
\ddot\chi=c^2\chi''-(\chi-\nu t)-\gamma\,\phi(\dot\chi).
\label{continuum}\end{equation}
$\chi(x,t)$ represents the time-dependent local longitudinal deformation of 
the surface of the upper plate in the static reference frame of the lower 
plate; dots and primes are temporal and spatial derivatives, respectively.
In this continuum limit, the number of blocks $N$ becomes the system size $S$.
$\phi(\dot\chi)$ is our normalized asymptotically velocity strengthening 
friction. There are three dimensionless parameters: $\gamma$ measures the 
magnitude of the friction, $c$ is the longitudinal speed of sound, 
and $\nu$ represents the pulling velocity or slip rate. From 
Eq.\ (\ref{continuum}), we obtain an expression for the local 
velocity $\psi=\dot\chi$ of the interface that, written as a couple of 
differential equations of first order in time, gives us our continuum 
Burridge--Knopoff model
\begin{mathletters}
\begin{eqnarray}
\dot\psi&=&\gamma(\eta-\phi(\psi)), 
\label{eq1} \\
\dot\eta&=&-\gamma^{-1}(\psi-\nu-c^2\psi'')
\label{eq2}
.\end{eqnarray}\label{eq}
\end{mathletters}
With the proper choice of the function $\phi(\psi)$, and if we may disregard 
for a moment the positioning of the spatial derivatives term, Eqs.\ (\ref{eq})
constitute a version \cite{fgpv} of the van der Pol--FitzHugh--Nagumo model 
that is the prototypical description of an excitable medium. 

While this type of medium was first studied in the physiology of cardiac 
and nervous tissues, the same properties are also seen in chemical and 
physical systems. An electrical caricature of an element of an excitable 
biological membrane is given by the circuit shown in Fig.\ \ref{circuit}(a). 
The membrane represented by the capacitor $C$ is charged by ion pumps 
characterized by the current generator $i_0$ and drained through a nonlinear 
resistance $n_l$ across the membrane. This nonlinear element should have the 
$v-i$ characteristic $\phi$ shown in Fig.\ \ref{circuit}(b), and could be 
implemented by a tunnel diode, 
a neon lamp, or any electronic device with a range of negative resistance.
The inductance $L$ models the finite switching time of the ion channels
in the membrane. The circuit equations are then formally equivalent to 
Eqs.\ (\ref{eq}) above, where $\psi$ and $\eta$ are proportional
to the current and potential difference across the nonlinear device, 
respectively, and the current $i_0$ is proportional to $\nu$.
Such circuit elements have been used in electronic experiments with 
Burridge--Knopoff models \cite{field}. In the models of excitable 
media of FitzHugh \& Nagumo, $\phi$ is generally taken to be as in 
Fig.\ \ref{circuit}(b): it thus has exactly the same form as used in our 
Burridge--Knopoff model (cf. Fig.\ \ref{friction}(b)).

\begin{figure}
\begin{center}
\def\epsfsize#1#2{0.46\textwidth}
\leavevmode
\epsffile{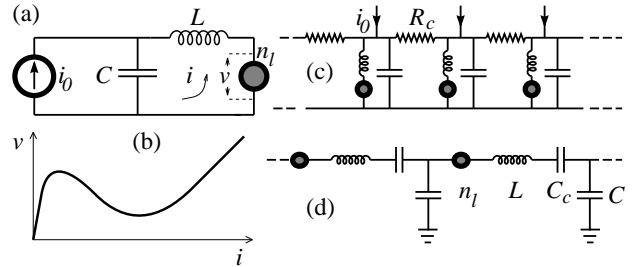}
\end{center}
\caption{\label{circuit}
An electronic excitable medium.
(a) An element of the medium: the circuit and (b) the $v-i$ characteristic of
$n_l$. (c) Resistive (diffusive) spatial coupling. (d) Capacitive (elastic) 
spatial coupling --- an active transmission line.
}
\end{figure}

The spatial derivatives in Eqs.\ (\ref{eq}) are unusual in the realm of
excitable media. In a biological excitable membrane, the excitation propagates
in space by diffusion. With the electrical model described above one can build
a spatially extended diffusive medium by coupling resistively several excitable
elements, as indicated in Fig.\ \ref{circuit}(c). Such a coupling would give
rise to current diffusion: the term involving $\psi''$ would appear in 
Eq.\ (\ref{eq1}) instead of in Eq.\ (\ref{eq2}). As they are written however, 
Eqs.\ (\ref{eq}) represent the network of capacitively coupled excitable 
elements illustrated in Fig.\ \ref{circuit}(d). This is a discrete 
representation of an active transmission line, where $L$ and $C$ are the 
distributed inductance and capacitance and $n_l$ is the distributed gain 
achieved by a nonlinear negative resistance along the line. The distributed 
serial capacitance $C_c$ represents the effects of dispersion in the line. The 
continuum limit of this network is a caricature of an active optical waveguide 
such as a fiber amplifier or laser.

Further motivation for the study of Eqs.\ (\ref{eq}) is provided by 
laboratory experiments \cite{rubio} in which an elastic gel was placed in a 
Taylor--Couette type apparatus. Stick--slip events occurred at the contact 
surface of the gel with the rotating inner cylinder. In a large part of the 
parameter space, low dimensional structures --- continuous slip, global 
oscillations, and propagating fronts --- were found rather than complex
spatiotemporal patterns that might have been expected to dominate. With this 
experiment in mind we focus on periodic boundary conditions. Since we are 
interested in qualitative aspects of the dynamics, we concentrate on the case 
where $\phi$ is taken to be a cubic polynomial. By shifting the origin of the 
variables $\psi$ and $\eta$ and the parameter $\nu$ we can reduce $\phi$ to 
the form $\phi(\psi)=\psi^3/3-\psi$. This makes our model a set of elastically
coupled van der Pol oscillators. For this election of $\phi$, the slip 
threshold, defined as the crossover as we increase $\nu$ from velocity 
strengthening creep to velocity weakening slip, lies at $\nu=-1$.

\begin{figure}
\begin{center}
\def\epsfsize#1#2{0.46\textwidth}
\leavevmode
\epsffile{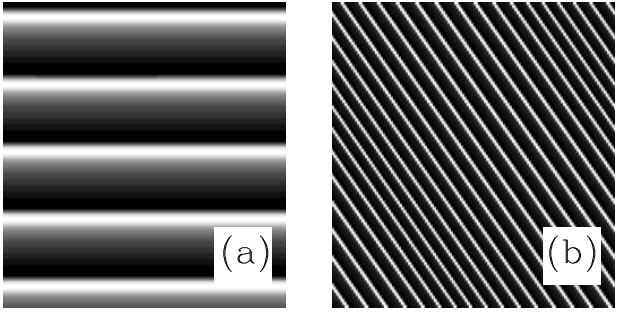}
\end{center}
\caption{\label{spattemp}
Spatiotemporal plots of $\psi(x,t)$ where time is vertical, space is 
horizontal. (a) A plot of the solution with stable uniform global 
oscillations for $\nu=-0.92$. (b) Stable propagating fronts at $\nu=-0.96$.
In both cases, the other parameter values are $\gamma=2$ and $c=0.3$. The
system size $S=20$, and the time elapsed $50$.
}
\end{figure}

The continuous slip solution, where one surface moves uniformly with respect 
to the other, was examined by Carlson \& Langer \cite{carlan}. They showed 
that the growth rate of perturbations of any wavelength at a particular slip 
rate is minus the slope of the friction function at that slip rate. This 
implies that with asymptotically velocity weakening friction the 
continuous slip solution is always unstable, while in our case 
this solution is stable for high and low slip rates where the slope is 
positive. Global oscillations, where one surface moves periodically in 
time with respect to the other, are also stable at some slip rates in our 
model. Figure \ref{spattemp}(a) shows the spatiotemporal pattern 
generated from an arbitrary initial condition with the system relatively 
far above the slip threshold. Here all points of space oscillate in phase 
with an instantaneous velocity that follows the dynamics of the van der Pol 
oscillator. To confirm the stability of these oscillations, we have 
calculated their Floquet multipliers. Global oscillations are solutions of 
Eqs.\ (\ref{eq}) with the spatial derivative set to zero, 
which correspond to the period-$T$ limit cycle of the van der Pol oscillator 
$\vec\Psi^*(t+T)=\vec\Psi^*(t)$ where $\vec\Psi^*(t)=(\psi^*(t),\eta^*(t))$, 
which is stable against spatially homogeneous
perturbations for $-1<\nu<1$ (i.e., above the slip threshold). 
We consider small spatial perturbations of the form 
$\vec\Psi^*(t)+\vec\varepsilon(t)\exp(iqx)$. 
Dropping quadratic and higher powers of 
$\vec\varepsilon(t)$ in Eqs.\ (\ref{eq}), 
the resulting linear equations
for $\vec\varepsilon(t)$ have time-periodic coefficients, 
and by the Floquet theorem their general solution is 
$\vec\varepsilon(t)=\vec\varepsilon_0 P(t)\exp(\Lambda_q t)$,
where $P(t)$ is a period-$T$ matrix and $\Lambda_q$ a time-independent
matrix. Once the limit cycle has been numerically determined, the matrix 
$\exp(\Lambda_q T)$ may be computed by numerically integrating the linearized
equations over the period $T$. Its two eigenvalues, the Floquet multipliers 
$(m_q^{(1)},m_q^{(2)})$, which may be either real, or complex conjugates, 
determine the stability of global oscillations under perturbations of 
wavenumber $q$. The growth rate per period of such a perturbation is the 
maximal Floquet exponent $\lambda_q=\max_i\{\Re[\ln(m_q^{(i)})]\}$. 
The dispersion relation $\lambda_q$ as a function of $q$ relatively far 
above the slip threshold is never positive, implying stability under 
perturbations of all scales. This contrasts with the same calculations 
performed for asymptotically velocity weakening friction, when 
there is always a range of long wavelengths for which perturbations grow 
exponentially, so that global oscillations are always unstable when this 
wavelength is within the system size. This is a significant difference between 
our model and Burridge--Knopoff models with asymptotically velocity weakening 
friction. Such global oscillations have been noted in laboratory friction 
experiments with elastic gels \cite{rubio}.

\begin{figure}
\begin{center}
\def\epsfsize#1#2{0.46\textwidth}
\leavevmode
\epsffile{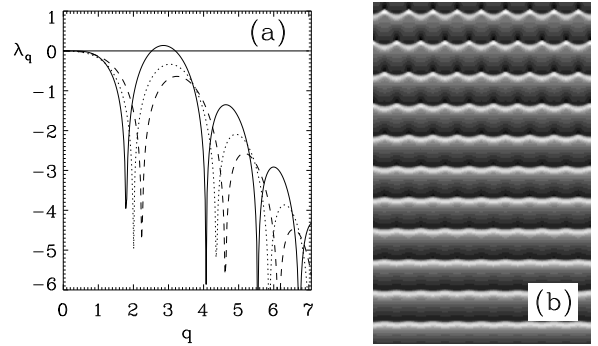}
\end{center}
\caption{\label{period2}
(a) A family of dispersion relations for the Floquet exponent showing the 
destabilization close to the slip threshold of global oscillations against 
perturbations of a single wavenumber $q_c\simeq3$. 
Shown are dispersion relations for 
(from the lower to the upper curve) $\nu=-0.90$, $-0.92$, and $-0.94$.
(b) Spatiotemporal plot of $\psi(x,t)$ for $\nu=-0.94$ showing the 
development of period-doubled structures resulting from this destabilization. 
Time elapsed is $125$, other settings are as in Fig.\ \ref{spattemp}. 
The spatial periodicity is determined by $q_c$.
}
\end{figure}

For slip rates not far above the slip threshold, global oscillations become 
unstable. The dispersion relation in Fig.\ \ref{period2}(a) shows that the 
destabilization occurs at a single finite wavenumber $q_c$. Furthermore,
the Floquet multiplier $m_{q_c}$ associated with the maximal Floquet exponent
crosses the complex unit circle through $-1$ at this point, indicating a 
period-doubling bifurcation. Figure \ref{period2}(b) is a spatiotemporal 
plot showing the nature of the structure arising near this instability. Small 
perturbations from the global uniformly oscillating regime grow to become
structures with a well-defined wavenumber $q_c$. For system size $S$, such 
structures evolve into a configuration of $q_c S/(2\pi)$ synchronized 
pacemakers emitting fronts in both directions. These fronts annihilate each 
other at the corresponding $q_c S/(2\pi)$ points, that then in turn become 
pacemakers. The resulting spatiotemporal structure is such that at an 
arbitrary point in space the periodicity in time is twice that of the 
original oscillations. These period-doubled structures are not well known 
in pattern formation, so their presence here is an interesting theoretical 
prediction. 

At slip rates right above the slip threshold, the front solutions cease 
to annihilate each other and evolve into fronts that propagate right around 
the system (Fig.\ \ref{spattemp}(b)). One can get an analytical handle on 
these propagating fronts by positing solutions of the type
$\psi(x,t)=f(\tilde z)$, where $\tilde z=x/v+t$, and $v$ is the 
front velocity. This ansatz with the further rescaling
$z=\tilde z/\sqrt{1-c^2/v^2}$ leads to 
\begin{equation}
\frac{d^2f}{dz^2}+\mu(f^2-1)\frac{df}{dz}+f=\nu
,\label{fronts}\end{equation}
which is again the van der Pol equation, but with the nonlinearity rescaled
by $\mu=\gamma/\sqrt{1-c^2/v^2}$. The propagating fronts are then periodic
solutions of the van der Pol equation. The parameter $\mu$ is undefined until 
the value of the front velocity $v$ is chosen. However, we know that the 
period of the solution is a function $T=T(\mu)$ of $\mu$: in the limit of 
large $\mu$, $T$ behaves as $T=k\mu+O(\mu^{-1})$, where 
$k=3+(\nu^2-1)\ln[(4-\nu^2)/(1-\nu^2)]$ \cite{fgpv}. 
Since this period should be commensurate with the system size $S$,
we have the condition $nT(\mu(v))=S/(v\sqrt{1-c^2/v^2})$, where $n$ is an
integer, to select the allowed front velocities, which in the large $\mu$ 
limit gives us the quantizing condition $v=S/(nk\gamma)$. Because 
Eq.\ (\ref{fronts}) has bounded solutions only if $v^2>c^2$, the propagating 
fronts are supersonic. Such supersonic propagating fronts have also been noted 
in Burridge--Knopoff models with asymptotically velocity weakening friction 
\cite{svr,lme}.

The original Burridge--Knopoff model was introduced as a means to reproduce 
the gross features of the statistics of real earthquakes \cite{burridge}. 
The Gutenberg--Richter power-law distributions \cite{gutenberg} in the 
statistics of slip events in the model have been considered as an example of 
self-organized criticality \cite{carlson}, and have been related to the 
presence of infinitely many degrees of freedom in the system \cite{cls}. 
However, recent numerical 
experiments \cite{svr,ricexu} indicate that power-law distributions of slip 
events may be due to discretization, finite size, and transient effects, and 
are not present at long times in the continuum limit. This has lead to a 
questioning of the relevance of the model to earthquakes in the real world. 
Our aim here, however, has been to investigate a Burridge--Knopoff model 
with a physically interesting friction law relevant to laboratory friction
experiments: ours are simply {\em laboratory earthquakes}.

We have returned to the asymptotically velocity strengthening type of friction 
law originally introduced by Burridge \& Knopoff, which was abandoned by 
Carlson \& Langer and later investigators, and with it have been successful in 
explaining some results of laboratory friction experiments. Our model 
constitutes a form of excitable medium, not previously considered, with elastic 
instead of diffusive coupling between spatial elements. We have studied the 
spatiotemporal dynamics of elastic excitable media and find novel dynamics. We 
believe that this type of elestic excitable medium may have applications 
beyond laboratory friction experiments to electronic transmission lines and 
active optical waveguides.

We should like to thank Miguel Angel Rubio and Ed Spiegel for useful 
discussions.
We acknowledge the financial support of the Spanish Direcci\'on General de 
Investigaci\'on Cient\'\i fica y T\'ecnica, contracts PB94-1167 and 
PB94-1172.

\end{twocolumns}

\begin{references}
\bibitem[*]{jemail}
julyan@hp1.uib.es, WWW http://formentor.uib.es/$\sim$julyan.
\bibitem[\dagger]{hemail}
dfsehg4@ps.uib.es, WWW http://formentor.uib.es/$\sim$emilio.
\bibitem[\ddagger]{oemail}
piro@hp1.uib.es, WWW http://formentor.uib.es/$\sim$piro.
\bibitem{vdpfn}
B. van~der Pol and J. van~der Mark, Phil. Mag. {\bf 6}, 763 (1928);
R. FitzHugh, J. Gen. Physiol. {\bf 43}, 867 (1960);
Biophys. J. {\bf 1}, 445 (1961);
J.~S. Nagumo, S. Arimoto, and S. Yoshizawa, Proc. IREE Aust. {\bf 50}, 2061
(1962).
\bibitem{meron}
E. Meron, Phys. Rep. {\bf 218}, 1 (1992).
\bibitem{burridge}
R. Burridge and L. Knopoff, Bull. Seismol. Soc. Am. {\bf 57}, 341 (1967).
\bibitem{carlson}
J.~M. Carlson and J.~S. Langer, Phys. Rev. Lett. {\bf 62}, 2632 (1989).
\bibitem{heslot}
F. Heslot {\it et~al.}, Phys. Rev. E {\bf 49}, 4973 (1994).
\bibitem{brechet}
Y. Brechet and Y. Estrin, Scripta Metall. {\bf 30}, 1449 (1994).
\bibitem{kilgore}
B.~D. Kilgore, M.~L. Blanpied, and J.~H. Dieterich, Geophys. Res. Lett. 
{\bf 20}, 903 (1993).
\bibitem{persson}
B.~N.~J. Persson, Phys. Rev. B {\bf 51}, 13568 (1995).
\bibitem{rubio}
M.~A. Rubio and J. Galeano, Phys. Rev. E {\bf 50}, 1000 (1994).
\bibitem{hong}
D.~C. Hong and S. Yue, Phys. Rev. Lett. {\bf 74}, 254 (1995).
\bibitem{kurtze}
D.~A. Kurtze and D.~C. Hong, Phys. Rev. Lett. {\bf 71}, 847 (1993).
\bibitem{kubleb}
L.~P. Kubin and Y. Estrin, Acta Metall. {\bf 33}, 397 (1985);
M.~A. Lebyodkin, Y. Brechet, Y. Estrin, and L.~P. Kubin, Phys. Rev. Lett. 
{\bf 74}, 4758 (1995).
\bibitem{carlan}
J.~M. Carlson and J.~S. Langer, Phys. Rev. A {\bf 40},  6470  (1989).
\bibitem{fgpv}
M. Feingold, D.~L. Gonz\'alez, O. Piro, and H. Viturro, Phys. Rev. A {\bf 37},
4060 (1988).
\bibitem{field}
S. Field, N. Venturi, and F. Nori, Phys. Rev. Lett. {\bf 74}, 74 (1995).
\bibitem{svr}
J. Schmittbuhl, J.-P. Vilotte, and S. Roux, Europhys. Lett. {\bf 21}, 375 
(1993).
\bibitem{lme}
J.~S. Langer and C. Tang, Phys. Rev. Lett. {\bf 67}, 1043 (1991);
C.~R. Myers and J.~S. Langer, Phys. Rev. E {\bf 47}, 3048 (1993);
P. Espa{\~n}ol, {\it ibid.\/} {\bf 50}, 227 (1994).
\bibitem{gutenberg}
B. Gutenberg and C.~F. Richter, Ann. Geophys. {\bf 9}, 1 (1956).
\bibitem{cls}
J.~M. Carlson, J.~S. Langer, and B.~E. Shaw, Rev. Mod. Phys. {\bf 66}, 657
(1994).
\bibitem{ricexu}
J.~R. Rice, J. Geophys. Res. {\bf 98}, 9885 (1993);
H.-J. Xu and L. Knopoff, Phys. Rev. E {\bf 50}, 3577 (1994).
\end{references}
\end{document}